\documentclass[12pt]{article}
\usepackage{url}
\usepackage{hyperref}
\usepackage{amsmath}
\usepackage{amsfonts}
\usepackage{amsthm}
\usepackage{graphicx}
\usepackage{epstopdf}
\usepackage{cite}
\usepackage[ruled,vlined]{algorithm2e}
\usepackage{enumitem}
\usepackage{siunitx}

\usepackage[top=0.7in, bottom=0.7in, left=0.7in, right=0.7in]{geometry}
\newcommand{\ud}{\,\mathrm{d}}
\title{LINOEP vectors, spiral of Theodorus, and nonlinear time-invariant system models of mode decomposition}
\author{Pushpendra Singh$^{1,2,}$\footnote{Corresponding author's E-mail address: \texttt{pushpendra.singh@ee.iitd.ernet.in; spushp@gmail.com} (P. Singh).  
$^1$\url{http://ee.iitd.ernet.in/} ; $^2$\url{www.jiit.ac.in/}}
\\{\normalsize $^{1}$Department of EE, Indian Institute of Technology Delhi, India}\\
{\normalsize $^{2}$Jaypee Institute of Information Technology - Noida, India}}
\providecommand{\keywords}[1]{\textbf{\textit{Keywords:}} #1}
\date{}
\begin{document}
\maketitle
\begin{abstract}
In this paper, we propose a general method to obtain a set of Linearly Independent Non-Orthogonal yet Energy (square of the norm) Preserving (LINOEP) vectors using iterative filtering operation and we refer it as Filter Mode Decomposition (FDM). We show that the general energy preserving theorem (EPT), which is valid for both linearly independent (orthogonal and nonorthogonal) and linearly dependent set of vectors, proposed by Singh P. et al. is a generalization of the discrete spiral of Theodorus (or square root spiral or Einstein spiral or Pythagorean spiral). From the EPT, we obtain the (2D) discrete spiral of Theodorus and show that the multidimensional discrete spirals (e.g. a 3D spiral) can be easily generated using a set of multidimensional energy preserving unit vectors.
We also establish that the recently proposed methods (e.g. Empirical Mode Decomposition (EMD), Synchrosqueezed Wavelet Transforms (SSWT), Variational Mode Decomposition (VMD), Eigenvalue Decomposition (EVD), Fourier Decomposition Method (FDM), etc.), for nonlinear and nonstationary time series analysis, are nonlinear time-invariant (NTI) system models of filtering. Simulation and numerical results demonstrate the efficacy of LINOEP vectors.
\end{abstract}
 \section{Introduction}
The Fourier introduced the trigonometric series to obtain the solutions of the heat equation, which is a diffusion partial differential equation (PDE), in a metal plate.
The Fourier representation has been applied to a wide range of physical and mathematical problems including electrical engineering, signal processing, image processing, vibration analysis, acoustics, optics, quantum mechanics, wave propagation, econometrics, etc.

A set of functions are called orthogonal if their inner product, other than itself, is zero.
Representation of a signal (or function) as a sum of series of orthogonal functions, motivated by Fourier series, is most important mathematical function expansion model for engineering systems and physical phenomena. The Parseval's theorem state that the energy in temporal space is same as energy in spectral space and hence energy is preserved in orthogonal series expansion of a signal.
The energy preserving property is important for a variety of reasons, and it is obtained by the decomposition of a signal into sum of orthogonal functions in various transforms like Fourier, Wavelet, Fourier-Bessel, spherical harmonics, Legendre polynomials, etc.

In the literature, there are various methods and applications \cite{rs1,rs13,rs14,FDM,TaylorsNP,PSEEG} of 1D nonlinear and nonstationary time series. The linearly independent (LI), non orthogonal yet energy (square of the norm) preserving (LINOEP) vectors are introduced in~\cite{rs19} and generated through energy preserving empirical mode decomposition (EPEMD) algorithm.
A set of linearly independent (LI) vectors can be transformed to a set of orthogonal vectors by the Gram-Schmidt Orthogonalization Method (GSOM). Similar to the GSOM, a LINOEP method is presented in~\cite{LINOEP} that transforms a set of LI vectors to a set of LINOEP vectors in an inner product space. It is also shown that there are many solutions to preserve the square of the norm in an inner product space.

We present general algorithms to obtain LINOEP vectors from the decomposition of a signal through iterative filtering operation. These algorithms use FILTER which can be linear (e.g. FIR, IIR), nonlinear (e.g. Empirical Mode Decomposition (EMD)~\cite{rs1}, Synchrosqueezed Wavelet Transforms (SSWT)~\cite{SWT}, Variational Mode Decomposition (VMD)~\cite{VMD}, Eigenvalue Decomposition (EVD)~\cite{EVD}, Fourier Decomposition Method (FDM)~\cite{FDM}, etc. ), time-invariant or time-variant filtering operation.

The classical discrete spiral of Theodorus (or square root spiral or Einstein spiral or Pythagorean spiral) can be constructed from the sequence of right triangles, with length of sides ($\sqrt{l},1, \sqrt{l+1}$) for $l=1,2,3,\cdots,17$ (originally and later extended for any $l\in \mathbb{N}$), which are arranged such that the first one has a cathetus on the real axis and all of triangles have the origin as a common vertex on the coordinate system of complex plane or 2D vector space (as shown in Figure~\ref{fig:dsot}). The discrete spiral of Theodorus, its mathematical properties and extension have been studied by several authors, e.g. \cite{soTh,soTh1,soTh2,soTh3,soTh4}, in detail. Here, we show that the general energy (square of the norm) preserving theorem (EPT) (valid for orthogonal, linearly independent but nonorthogonal, and linearly dependent vectors) proposed in~\cite{rs19} is a generalization of the discrete spiral of Theodorus. This EPT has same special structure in multidimensional (ND) vector space which is present in the discrete spiral of Theodorus in 2D.

This paper is organized as follows: the orthogonal and LINOEP vectors are discussed in Section 2. Filter Mode Decomposition: A general approach to obtain LINOEP vectors and nonlinear time-invariant (NTI) systems model of filtering for mode decomposition of signals are discussed in Section 3. Simulation and numerical results are given in Section 4. Section 5 presents conclusions.
\section{The orthogonal and LINOEP vectors}
In this section, we present a brief overview of the orthogonal and LINOEP vectors in the Hilbert space over the field of complex number as follows.
\theoremstyle{definition}
\newtheorem{defn}{Definition}[section]
\begin{defn}
Let $H$ be a Hilbert space.
\begin{enumerate}
\item Two vectors $\mathbf{f}, \mathbf{g}\in H$ are orthogonal if inner product $\langle \mathbf{f,g}\rangle=0$.
\item A sequence of vectors $\{\mathbf{f}_k\}_{k=1}^n \in H$ is an orthogonal sequence if $\langle \mathbf{f}_k,\mathbf{f}_l\rangle=0$ for $k \ne l$.
\item A sequence of vectors $\{\mathbf{e}_k\}_{k=1}^n\in H$ is an orthonormal sequence if it is orthogonal and each vector is a unit vector, i.e. $\langle \mathbf{e}_k,\mathbf{e}_l\rangle=1$ for $k = l$, and $\langle \mathbf{e}_k,\mathbf{e}_l\rangle=0$ for $k \ne l$.
\end{enumerate}
\end{defn}
\newtheorem{mydef}{Theorem}[section]
\begin{mydef}[Pythagorean Theorem] \label{PTThrm}
If $\mathbf{f}_1,  \cdots, \mathbf{f}_n \in H$ are orthogonal and $\mathbf{x}=\sum_{k=1}^{n}\mathbf{f}_k$, then
\begin{equation}
\left\lVert \mathbf{x} \right\rVert^2=\left\lVert \sum_{k=1}^{n}\mathbf{f}_k \right\rVert^2 =\sum_{k=1}^{n}\lVert \mathbf{f}_k\rVert^2. \label{PTeqn}
\end{equation}
\end{mydef}
The LI nonorthogonal yet energy preserving (LINOEP) class of vectors and the following theorem are proposed, in~\cite{rs19}, for the development of EPEMD algorithm.
\begin{mydef}[Energy Preserving Theorem] \label{LINOEPThrm}
Let $H$ be a Hilbert space over the field of complex numbers, and let $\{\mathbf{x},\mathbf{x}_1,\cdots,\mathbf{x}_{n}\}$ be a set of vectors satisfying the following conditions:
\begin{equation}
  {\mathbf{x}_k \perp \sum_{l=k+1}^{n}\mathbf{x}_l}, \qquad \text{ for } k=1,2,\cdots, n-1, \label{st1}
\end{equation}
\begin{equation}
\text{ and, } \quad \mathbf{x} =\sum_{k=1}^{n}\mathbf{x}_k. \label{st2}
\end{equation}
Then in the representation, given in \eqref{st2}, the square of the norm, and hence energy is preserved, i.e.
\begin{equation}
\left\lVert \mathbf{x} \right\rVert^2= \left\lVert \sum_{k=1}^{n}\mathbf{x}_k \right\rVert^2 =\sum_{k=1}^{n}\lVert \mathbf{x}_k\rVert^2.
\end{equation}
\end{mydef}
\noindent We observe three interesting cases of this energy preserving theorem:
\begin{enumerate}
\item If set of vectors $\{\mathbf{x}_1,\cdots,\mathbf{x}_{n}\}$ is orthogonal set, then this is Pythagorean Theorem~\ref{PTThrm}.
\item If set of vectors $\{\mathbf{x}_1,\cdots,\mathbf{x}_{n}\}$ is linearly independent (LI) but nonorthogonal set, then we refer this as LINOEP Theorem.
In this case, pairwise orthogonality is not required and only last two vectors (i.e. $\mathbf{x}_{n-1}$ and $\mathbf{x}_n$) are orthogonal.
It is interesting to note that, although vectors in the LINOEP theorem are not orthogonal, yet it satisfy the same property (preserve the energy or square of the norm) that is being satisfied the Pythagorean's theorem when all vector are orthogonal.
\item If set of vectors $\{\mathbf{x}_1,\cdots,\mathbf{x}_{n}\}$ is linearly dependent (LD) set (e.g. if only $k$ vectors are LI, then $n-k$ vectors are LD), then this is simple energy preserving Theorem. We observe that the discrete spiral of Theodorus can be obtained from this case (see Section~\ref{dsTh}).
\end{enumerate}
Using the above discussions, we present the following definition of a sequence of energy preserving vectors.
\begin{defn}[energy preserving vectors]
Let $H$ be a Hilbert space. A sequence of vectors $\mathbf{x}_1,\mathbf{x}_2,\cdots,\mathbf{x}_n\in H$ is an energy preserving sequence if $\langle \mathbf{x}_k , \sum_{l=k+1}^{n}\mathbf{x}_l \rangle=0$ for $k=1,2,\cdots, n-1$.
\end{defn}
Now, we consider the following examples of inner products and the norm induced by these inner products, which would be used for calculation of numerical values (of energy and percentage error) in simulation results.
\\\textbf{Examples:} 
(a) For time or spatial series (1D discrete signal) $\mathbf{f}=[f_1,\cdots,f_n]^T \in \mathbb{C}^n$ and $\mathbf{g}=[g_1,\cdots,g_n]^T \in \mathbb{C}^n$, inner product can be defined as
\begin{equation}
\langle \mathbf{f,g}\rangle=\sum_{i=1}^n f_i\bar{g}_i, \label{IPeqn1}
\end{equation}
and the norm induced by the inner product \eqref{IPeqn1} is defined as
\begin{equation}
\left\lVert \mathbf{f} \right\rVert=\sqrt{\langle \mathbf{f,f}\rangle}=\sqrt{\sum_{i=1}^n |f_i|^2}. \label{IPNeqn1}
\end{equation}
(b) For image signal (2D discrete signal) $\mathbf{x}$ and $\mathbf{y} \in \mathbb{C}^{m \times n}$, inner product can be defined as
\begin{equation}
\langle \mathbf{x,y}\rangle=\sum_{k=1}^m\sum_{l=1}^n x_{kl}\bar{y}_{kl}, \label{IPeqn2}
\end{equation}
and the norm induced by the inner product \eqref{IPeqn2} is defined as
\begin{equation}
\left\lVert \mathbf{x} \right\rVert=\sqrt{\langle \mathbf{x,x}\rangle}=\sqrt{\sum_{k=1}^m\sum_{l=1}^n |x_{kl}|^2}. \label{IPNeqn2}
\end{equation}
The energy of a signal is defined as square of the norm, i.e. $E_\mathbf{f}=\left\lVert \mathbf{f} \right\rVert^2={\langle \mathbf{f,f}\rangle}$ and $E_\mathbf{x}=\left\lVert \mathbf{x} \right\rVert^2={\langle \mathbf{x,x}\rangle}$ for 1D and 2D discrete signals, respectively.
\section{Filter Mode Decomposition}
In this section, first, we propose a general approach to obtain LINOEP vectors by a algorithm which we refer as Filter Mode Decomposition (FMD) and show that the decomposition of a signal into a set of LINOEP vectors is more natural than set of orthogonal vectors by non ideal filtering methods. A filter is ideal if it has a brick wall (rectangular) frequency response.
Second, we show that the nonlinear and nonstationary time series decomposition and analysis methods, such as EMD, SSWT, VMD, EVD, FDM, etc. are nonlinear time-invariant (NTI) system models of iterative filtering, which belong to general class of FMD algorithm.
\subsection{A general approach to obtain LINOEP vectors}
Here, we propose a general approach to obtain LINOEP vectors using iterative filtering operations.
Filtering operation is a class of signal processing that transfers the desired frequency components and removes completely or partially unwanted frequency components or features of a signal. Usually the filters are of low pass, high pass, band pass or band stop type. Filters may be analog or digital, linear or nonlinear, time-invariant or time-variant, discrete-time or continuous-time, linear-phase or nonlinear-phase, infinite impulse response (IIR) or finite impulse response (FIR) type of discrete-time filter, etc.

A linear time-invariant (LTI) zero-phase filter is a special case of a linear-phase filter in which the phase slope of frequency response is zero, i.e. its frequency response, always greater than zero in the filter passbands, is a real and even function of frequency. A LTI zero-phase filter cannot be causal except in the trivial case when the filter operation is a constant scale multiplier to a signal. However, in many off-line applications, such as when filtering a data file stored on a memory, causality is not a requirement, and zero-phase filters are preferred because zero-phase filtering preserves salient features (e.g. maxima, minima, etc.) in the filtered time waveform exactly at the time where those features occur in the unfiltered waveform.

General filter mode decomposition algorithms are presented in Algorithm 1, Algorithm 2 and Algorithm 3. Algorithm 2 and Algorithm 3 present general approach to obtain LINOEP vectors using iterative filtering operations. Both algorithms decompose a signal into its LINOEP components which follow the mathematical model of Theorem \ref{LINOEPThrm}. In Algorithm 2, second vector, after each stage of filtering, is taken as one plus constant multiplier of residue, i.e. $\mathbf{\tilde{c}}_{i+1}= (1+\alpha_i) \mathbf{r}_i$, and first vector is obtained by subtracting the constant multiplier of residue form a filter output i.e. $\mathbf{c}_i= \mathbf{y}_i- \alpha_i \mathbf{r}_i$. The value of $\alpha_i$ is obtained such that vectors $\mathbf{c}_i$ and $\mathbf{\tilde{c}}_{i+1}$ are orthogonal, i.e. ${\langle \mathbf{c}_i,  \mathbf{\tilde{c}}_{i+1} \rangle}=0$. For each iteration, sum of signal and its energy is preserved, i.e. $\mathbf{x}_i=\mathbf{r}_i+\mathbf{y}_i=\mathbf{c}_i+\mathbf{\tilde{c}}_{i+1}$ and $\lVert\mathbf{x}_i\rVert^2=\lVert\mathbf{c}_i\lVert^2+\lVert\mathbf{\tilde{c}}_{i+1}\rVert^2$.

In Algorithm 3, first vector, after each stage of filtering, is taken as one plus constant multiplier of residue, i.e. $\mathbf{v}_{i}= (1+\alpha_i) \mathbf{y}_i$, and second vector is obtained by subtracting the constant multiplier of residue form a filter output i.e. $\mathbf{\tilde{v}}_{i+1}= \mathbf{r}_i- \alpha_i \mathbf{y}_i$. The value of $\alpha_i$ is obtained such that vectors $\mathbf{v}_i$ and $\mathbf{\tilde{v}}_{i+1}$ are orthogonal, i.e. ${\langle \mathbf{v}_i,  \mathbf{\tilde{v}}_{i+1} \rangle}=0$. For each iteration, sum of signal and its energy is preserved, i.e. $\mathbf{x}_i=\mathbf{r}_i+\mathbf{y}_i=\mathbf{v}_i+\mathbf{\tilde{v}}_{i+1}$ and $\lVert\mathbf{x}_i\rVert^2=\lVert\mathbf{v}_i\lVert^2+\lVert\mathbf{\tilde{v}}_{i+1}\rVert^2$.

\begin{algorithm}[!t]
 $\mathbf{x}_{1}=\mathbf{x}$\;
 \For{$i=1$ to $n$}
 {
  $\mathbf{y}_i=FILTER_i(\mathbf{x}_i)$\;
  $\mathbf{r}_i=\mathbf{x}_i-\mathbf{y}_i$\;
  $\mathbf{x}_{i+1}=\mathbf{r}_{i}$\;
 }
 $\mathbf{y}_{n+1}=\mathbf{r}_{n}$\;
 \caption{A FMD algorithm to obtain vectors $\mathbf{y}_i$ from decomposition of a signal $\mathbf{x}$ such that $\mathbf{x}=\sum_{i=1}^{n+1}\mathbf{y}_i$. It is to be noted that, in general, set $\{\mathbf{y}_i,\cdots, \mathbf{y}_{n+1}\}$ is neither orthogonal set nor LINOEP one, hence energy is not preserved in decomposition.}
\end{algorithm}\label{linoep_algo0}
\begin{algorithm}[!t]
 $\mathbf{x}_{1}=\mathbf{x}$\;
 \For{$i=1$ to $n$}
 {
  $\mathbf{y}_i=FILTER_i(\mathbf{x}_i)$\;
  $\mathbf{r}_i=\mathbf{x}_i-\mathbf{y}_i$\;
  ${\alpha_i=\frac{\langle \mathbf{y}_i,\mathbf{r}_i \rangle }{\langle \mathbf{r}_i,\mathbf{r}_i \rangle }}$\;
  $\mathbf{c}_i= \mathbf{y}_i- \alpha_i \mathbf{r}_i$\;
  $\mathbf{\tilde{c}}_{i+1}= (1+\alpha_i) \mathbf{r}_i$\;
  $\mathbf{x}_{i+1}=\mathbf{\tilde{c}}_{i+1}$\;
 }
 $\mathbf{c}_{n+1}=\mathbf{\tilde{c}}_{n+1}$\;
 \caption{A FMD algorithm to obtain LINOEP vectors $\mathbf{c}_i$ from decomposition of a signal $\mathbf{x}$ such that $\mathbf{x}=\sum_{i=1}^{n+1}\mathbf{c}_i$ and ${\mathbf{c}_i \perp \sum_{l=i+1}^{n+1}\mathbf{c}_l}$. Values of $\alpha_i$ are computed such that $\mathbf{c}_i \perp \mathbf{\tilde{c}}_{i+1}$ for each iteration. It is to be noted that, in general, filter is not ideal (or non brick wall frequency response) one and hence ${\mathbf{c}_i} \not\perp {\mathbf{c}_l}$ for ${i,l=1,2,\dots,n}$ and only ${\mathbf{c}_n \perp \mathbf{c}_{n+1}}$.}
\end{algorithm}\label{linoep_algo}
\begin{algorithm}[!h]
 $\mathbf{x}_{1}=\mathbf{x}$\;
 \For{$i=1$ to $n$}
 {
  $\mathbf{y}_i=FILTER_i(\mathbf{x}_i)$\;
  $\mathbf{r}_i=\mathbf{x}_i-\mathbf{y}_i$\;
  ${\alpha_i=\frac{\langle \mathbf{y}_i,\mathbf{r}_i \rangle }{\langle \mathbf{y}_i,\mathbf{y}_i \rangle }}$\;
  $\mathbf{v}_i= (1+\alpha_i) \mathbf{y}_i$\;
  $\mathbf{\tilde{v}}_{i+1}=  \mathbf{r}_i - \alpha_i \mathbf{y}_i$\;
  $\mathbf{x}_{i+1}=\mathbf{\tilde{v}}_{i+1}$\;
 }
 $\mathbf{v}_{n+1}=\mathbf{\tilde{v}}_{n+1}$\;
 \caption{A FMD algorithm to obtain LINOEP vectors $\mathbf{v}_i$ from decomposition of a signal $\mathbf{x}$ such that $\mathbf{x}=\sum_{i=1}^{n+1}\mathbf{v}_i$ and ${\mathbf{v}_i \perp \sum_{l=i+1}^{n+1}\mathbf{v}_l}$. Values of $\alpha_i$ are computed such that $\mathbf{v}_i \perp \mathbf{\tilde{v}}_{i+1}$ for each iteration. It is to be noted that, in general, filter is not ideal one and hence ${\mathbf{v}_i} \not\perp {\mathbf{v}_l}$ for ${i,l=1,2,\dots,n}$ and only ${\mathbf{v}_n \perp \mathbf{v}_{n+1}}$.}
\end{algorithm}\label{linoep_algo1}
\subsection{Nonlinear time-invariant (NTI) systems model of filtering for mode decomposition of signals}
A system $S$ is time-invariant if its response to inputs or behavior does not change with time, i.e.
\begin{equation}
 S[x(t)]=y(t) \text{ and } S[x(t-\tau)]=y(t-\tau), \label{nti0}
\end{equation}
where, $y(t)$ is a output (response) to any input $x(t)$ to a system $S$ and $\tau$ is a delay parameter.
A system $S$ is linear if it follows the principle of superposition, which is combination of two properties: homogeneity (scaling) and additivity, i.e. for any $n$ signals $\{\mathbf{x}_l\}_{l=1}^n$ and any $n$ scalars $\{a_l\}_{l=1}^n$,
\begin{equation}
 S\Big[\sum_{l=1}^n a_l\mathbf{x}_l\Big]=\sum_{l=1}^n a_lS[\mathbf{x}_l]. \label{nti}
\end{equation}
In words, linearity means scaling and summing before or after the system are the same for all the input to output signal mappings. If a system $S$ is not following the the principle of superposition, then it is a nonlinear system, i.e. $ S\Big[\sum_{l=1}^n a_l\mathbf{x}_l\Big] \ne \sum_{l=1}^n a_lS[\mathbf{x}_l]$.

The EMD algorithm~\cite{rs1} can decompose a time series $x(t)$ into a set of finite band-limited IMFs and residue, i.e. $x(t) \to{EMD}\mapsto \{y_1(t),...,y_n(t),r_n(t)\}$ (or $EMD[x(t)] = \{y_1(t),...,y_n(t),r_n(t)\}$), such that the decomposed signal $x(t)$ is the sum of IMF components $\{y_i(t)\}_{i=1}^n$ and final residue $r_n(t)$:
\begin{equation}
 x(t)=\sum_{k=1}^{n}{y}_{k}(t) + r_n(t) =\sum_{k=1}^{n+1}{y}_{k}(t), \label{emd1}
\end{equation}
where $y_{k}(t)$ is the $k^{th}$ IMF and $r_n(t)=y_{n+1}(t)$. A set of IMFs obtained by EMD is neither orthogonal nor LINOEP vectors and, hence, energy of a signal is not preserved in decomposition. The energy preserving EMD (EPEMD) algorithm~\cite{rs19} decomposes a time series $x(t)$ into a set of finite band-limited IMFs and residue which follow the LINOEP vector model, i.e.
\begin{equation}
 x(t)=\sum_{k=1}^{n+1}{y}_{k}(t) \qquad \text{and} \qquad {{y}_{k}(t) \perp \sum_{l=k+1}^{n}{y}_{l}(t)}. \label{epemd1}
\end{equation}

All IMFs must satisfy two basic conditions~\cite{rs1}: (1) In the complete range of time series, the number of extrema (i.e. maxima and minima) and
the number of zero crossings are equal or differ at most by one. (2) At any point of time, in the complete range of time series,
the average of the values of upper and lower envelopes, obtained by the interpolation of local maxima and the local minima, is zero.

We observe that the EMD is following a nonlinear time-invariant (NTI) system model, and hence it is a iterative nonlinear time-invariant zero-phase filtering operations to decompose a signal into intrinsic mode functions (IMFs). All variants of the EMD algorithm are nonlinear because they don't follow the principle of superposition, i.e. there exist $n$ signals $\{\mathbf{x}_l\}_{l=1}^n$:
\begin{equation}
 EMD\Big[\sum_{l=1}^n \mathbf{x}_l\Big] \ne \sum_{l=1}^n EMD[\mathbf{x}_l]. \label{ntiEMD}
\end{equation}
Although it is easy to observe that EMD is a nonlinear system model, yet we provide counterexamples to prove this fact.

\textbf{Example 1:} Let $x_1(t)=\sin(10\pi t)$ and $x_2(t)=\sin(100\pi t)$ in time interval $0 \le t\le 1$ s. Clearly, both signal $x_1(t)$ and $x_2(t)$ are IMFs, ideally $EMD[x_1(t)]=x_1(t)$ and $EMD[x_2(t)]=x_2(t)$, thus $EMD[x_1(t)]+EMD[x_2(t)]=x_1(t)+x_2(t)$. However, it is very easy to verify, by the all variants of EMD algorithms, that ideally $EMD[x_1(t)+x_2(t)]=\{x_{1}(t), x_{2}(t)\}$ which implies that $EMD[x_1(t)+x_2(t)]\ne EMD[x_1(t)]+EMD[x_2(t)]$, i.e. $\{x_{1}(t), x_{2}(t)\} \ne \{x_1(t)+x_2(t)\}$. Clearly, EMD is a time-invariant system model as it follows~\eqref{nti0}, i.e. $EMD[x_1(t-\tau)]=x_1(t-\tau)$, $EMD[x_2(t-\tau)]=x_2(t-\tau)$ and $EMD[x_1(t-\tau)+x_2(t-\tau)]=\{x_{1}(t-\tau), x_{2}(t-\tau)\}$.

\textbf{Example 2:} Let $x_l(t)=(21-l)\sin(2\pi l f_0 t)$ for $l=1,2,\cdots, 20$, in time interval, $0 \le t\le 1$ s, with $T_0=1=\frac{1}{f_0}$. Clearly, all the $x_l(t)$ are IMFs and ideally $EMD[x_l(t)]=x_l(t)$. However, it is very easy to verify, by the all variants of EMD algorithms, that $EMD\Big[\sum_{l=1}^n x_l(t)\Big] \ne \sum_{l=1}^n EMD[x_l(t)]$, because EMD generates finite number of band-limited IMFs from the sum of all signals considered in this example.

It is very interesting to note that above arguments along with equation \eqref{ntiEMD} and Examples 1 and 2 are valid for the other nonlinear and nonstationary time series analysis methods, such as synchrosqueezed wavelet transforms, variational mode decomposition, eigenvalue decomposition, Fourier decomposition methods~\cite{FDM,FDM2D}, etc.

The FDM, entirely Fourier theory based decomposition, is recently proposed method~\cite{FDM} for the nonlinear and nonstationary time series analysis. The FDM algorithm can decompose a time series $x(t)$ into a set of finite band-limited Fourier intrinsic band functions (FIBFs) and constant, i.e. $x(t) \to{FDM}\mapsto \{y_1(t),...,y_n(t), c\}$ (or $FDM[x(t)] = \{y_1(t),...,y_n(t), c\}$), such that the decomposed signal $x(t)$ is the sum of orthogonal FIBF components $\{y_i(t)\}_{i=1}^n$ and a constant $c$:
\begin{equation}
 x(t)=\sum_{k=1}^{n}{y}_{k}(t) + c \qquad \text{and } \qquad \langle {y}_{k},c  \rangle = \langle {y}_{k},{y}_{l}  \rangle= 0 \text{ for } k\ne l, \label{fdm1}
\end{equation}
where $y_{k}(t)$ is the $k^{th}$ FIBF.
The FIBFs, $y_k(t) \in C^{\infty}[a,b]$, are functions that satisfy the following conditions~\cite{FDM}:
\begin{enumerate}
\item The FIBFs are zero mean functions, i.e. $\int_a^b y_k(t)\ud t=0$.
\item The FIBFs are orthogonal functions, i.e. $\int_a^b y_k(t) y_l(t)\ud t=0$, for $k\ne l$.
\item The FIBFs provide analytic FIBFs (AFIBFs) with instantaneous frequency (IF) and amplitude always greater than zero, i.e. $y_k(t)+j\hat{y}_k(t)=a_k(t)\exp(j\phi_k(t))$, with $a_k(t), \frac{\ud}{\ud t} \phi_k(t) \ge0$, $\forall t$.
\end{enumerate}
Where, $\hat{y}_k(t)$ is the Hilbert transform (HT) of FIBF $y_k(t)$, defined as convolution of $y_k(t)$ and $1/\pi t$, i.e. $\hat{y}_k(t)=y_k(t)*\frac{1}{\pi t}=\frac{1}{\pi} \text{ p.v.} \int_{-\infty}^{\infty}\frac{y_k(\tau)}{t-\tau} \ud\tau $, where p.v. stands for the Cauchy principal value of the integral. Even though the Hilbert transform is global, it emphasizes the properties of the function at the local time $t$. Thus, the HT is used to examine and reveal the local properties of the function $y_k(t)$ and hence $x(t)$ in \eqref{fdm1}.

The FDM algorithm directly generate orthogonal vectors, whereas for other nonlinear and nonstationary time series analysis methods (e.g. EMD, SSWT, EVD, etc.), it is more natural to generate LINOEP vectors as shown in Algorithm 2 and 3. From Algorithm 1, of course, one can generate  $(n+1)!$ sets of orthogonal vectors form a set of $(n+1)$ LI vectors, $\{\mathbf{y}_i,\cdots, \mathbf{y}_{n+1}\}$, by using the GSOM (as a post processing).
\section{Simulation and numerical results}
In this section, we present the discrete spiral of Theodorus and some simulation results which provide interesting properties of LINOEP vectors which have been obtained form linear time-invariant (LTI) zero-phase filtering operation on a signal.
\subsection{The discrete spiral of Theodorus}\label{dsTh}
\begin{figure}[!t]
\centering
\includegraphics[angle=0,width=0.7\textwidth,height=0.5\textwidth]{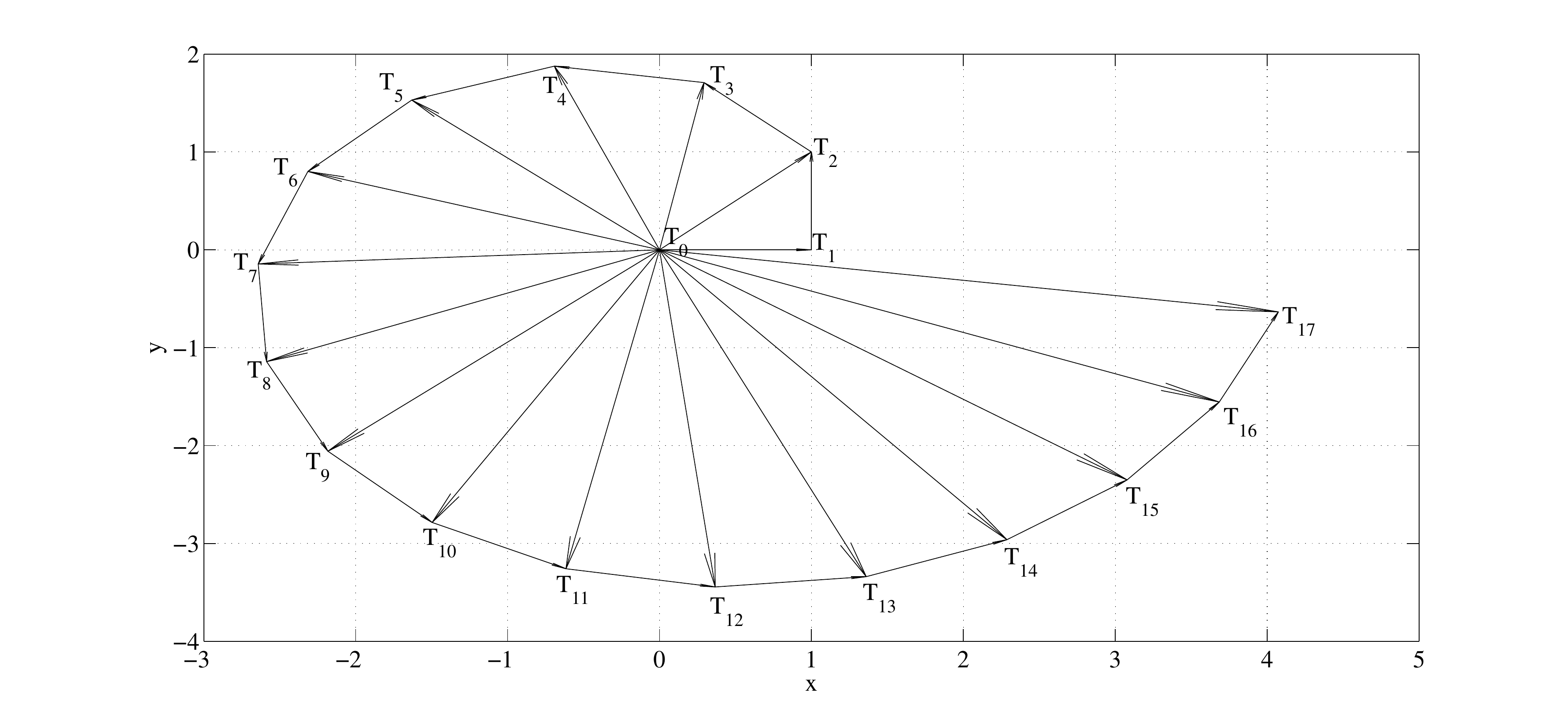}
\includegraphics[angle=0,width=0.7\textwidth,height=0.5\textwidth]{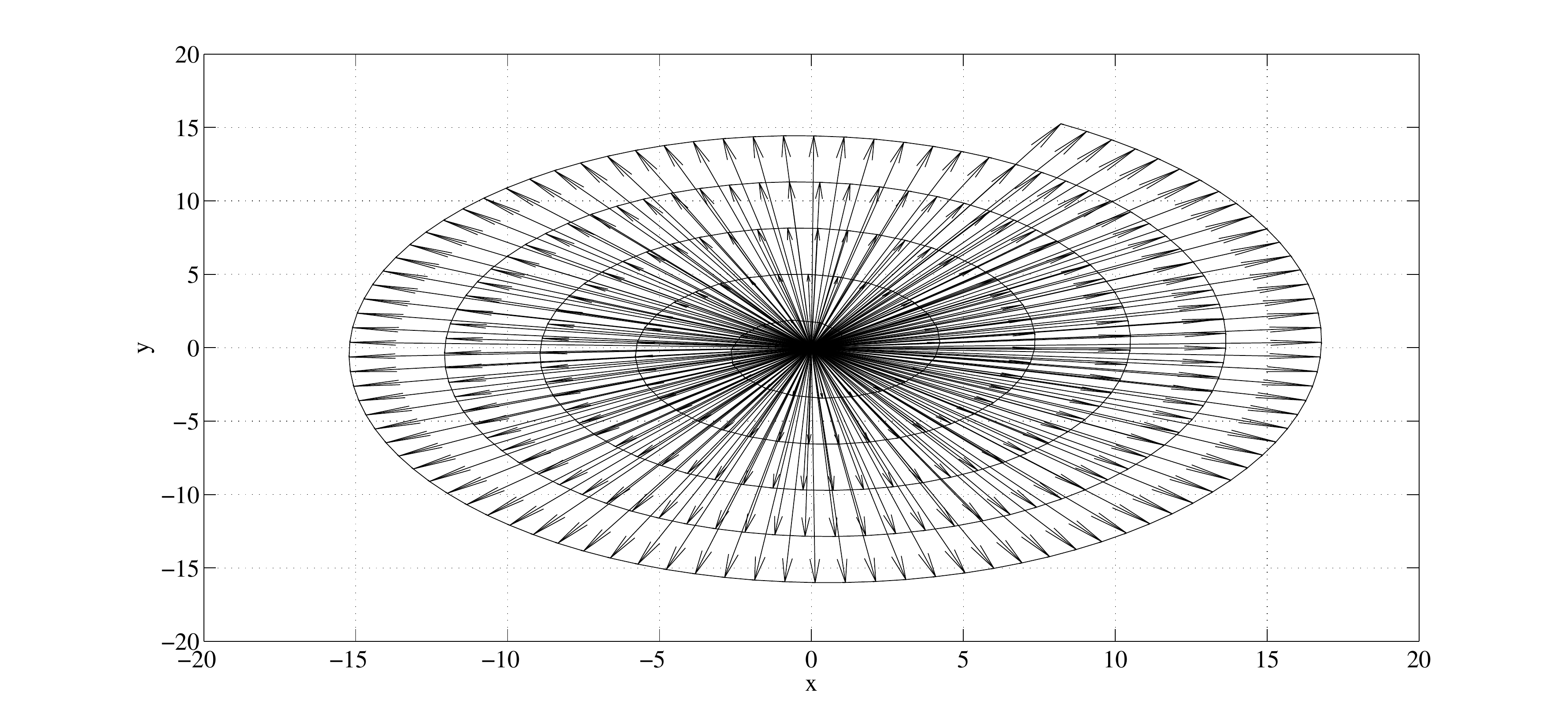} 
\caption{Discrete spiral of Theodorus in 2D (top: original one and drawn without any obstruction or overlap) $\left\lVert \mathbf{T}_{17} \right\rVert=\sqrt{17}$ and (bottom: extended version and drawn with intersections in figure) $\left\lVert \mathbf{T}_{300} \right\rVert=\sqrt{300}$.\label{fig:dsot}}
\end{figure}

We rewrite \eqref{st1} and \eqref{st2} as
\begin{equation}
  \sum_{l=1}^{k-1}\mathbf{x}_{l} \perp \mathbf{x}_k, \text{ for } k=2,3,\cdots,n \text{ and } \mathbf{x} =\sum_{k=1}^{n}\mathbf{x}_k, \qquad n\in \mathbb{N}.\label{dst1}
\end{equation}
Then from the EPT Theorem~\ref{LINOEPThrm}, we obtain
\begin{equation}
\left\lVert \mathbf{x} \right\rVert^2= \left\lVert \sum_{k=1}^{n}\mathbf{x}_k \right\rVert^2 =\sum_{k=1}^{n}\lVert \mathbf{x}_k\rVert^2. \label{dst3}
\end{equation}
Now, if we take a set of 2D unit vectors $\{\mathbf{x}_1,\cdots,\mathbf{x}_{n}\}$ which satisfy~\eqref{dst1} (which means only first two vectors are orthonormal vectors, i.e. $\mathbf{x}_{1} \perp \mathbf{x}_2$, and rest $(n-2)$ are LD unit vectors, i.e. they are obtained by the linear combinations of first two vectors), then we can define
\begin{equation}
\mathbf{T}_l=\sum_{k=1}^{l}\mathbf{x}_{k}, \quad l=1,2,\cdots,n. \label{dst31}
\end{equation}
From \eqref{dst1} and \eqref{dst31}, one can easily show that
 \begin{equation}
  \qquad \mathbf{T}_{l-1}\mathbf{T}_l=\mathbf{T}_l-\mathbf{T}_{l-1}=\mathbf{x}_{l}, \quad \mathbf{T}_{1}=\mathbf{x}_{1}, \quad \mathbf{T}_{l} \perp \mathbf{x}_{l+1}, \qquad  l=1,2,\cdots, n;\label{dst4}
\end{equation}
where $\mathbf{T}_{0}$ is the origin (zero vector).
The discrete spiral of Theodorus is shown in Firure~\ref{fig:dsot}, where $\mathbf{T_0}=[0 \quad 0]^T$. Let $\Phi_l$ is angle between $\mathbf{T}_1$ and $\mathbf{T}_l$, then from Figure~\ref{fig:dsot}, we obtain angle between $\mathbf{T}_{l+1}$ and $\mathbf{T}_l$ as
\begin{equation}
\tan(\Phi_{l+1}-\Phi_{l})=\frac{1}{\sqrt{l}} \Leftrightarrow (\Phi_{l+1}-\Phi_{l})=\tan^{-1}\Big(\frac{1}{\sqrt{l}}\Big), \qquad \Phi_1=0, \qquad l=1,2,\cdots, n. \label{phir_1}
\end{equation}
From above discussions and Figure~\ref{fig:dsot}, one can easily obtain
\begin{equation}
 \mathbf{T}_{l}=\sqrt{l}[\cos(\Phi_l)\quad \sin(\Phi_l)]^T, \quad \mathbf{T}_{1}=\mathbf{x}_{1}=[1 \quad 0]^T, \quad  \mathbf{x}_{l+1}=[-\sin(\Phi_l)\quad \cos(\Phi_l)]^T. \label{dst5}
\end{equation}
and hence $\left\lVert \mathbf{T}_l \right\rVert=\sqrt{l}$ and $\left\lVert \mathbf{x}_{l} \right\rVert=1$ (since $\mathbf{x}_{l}$ are unit vectors).

\begin{figure}[!t]
\centering
\includegraphics[angle=0,width=1\textwidth]{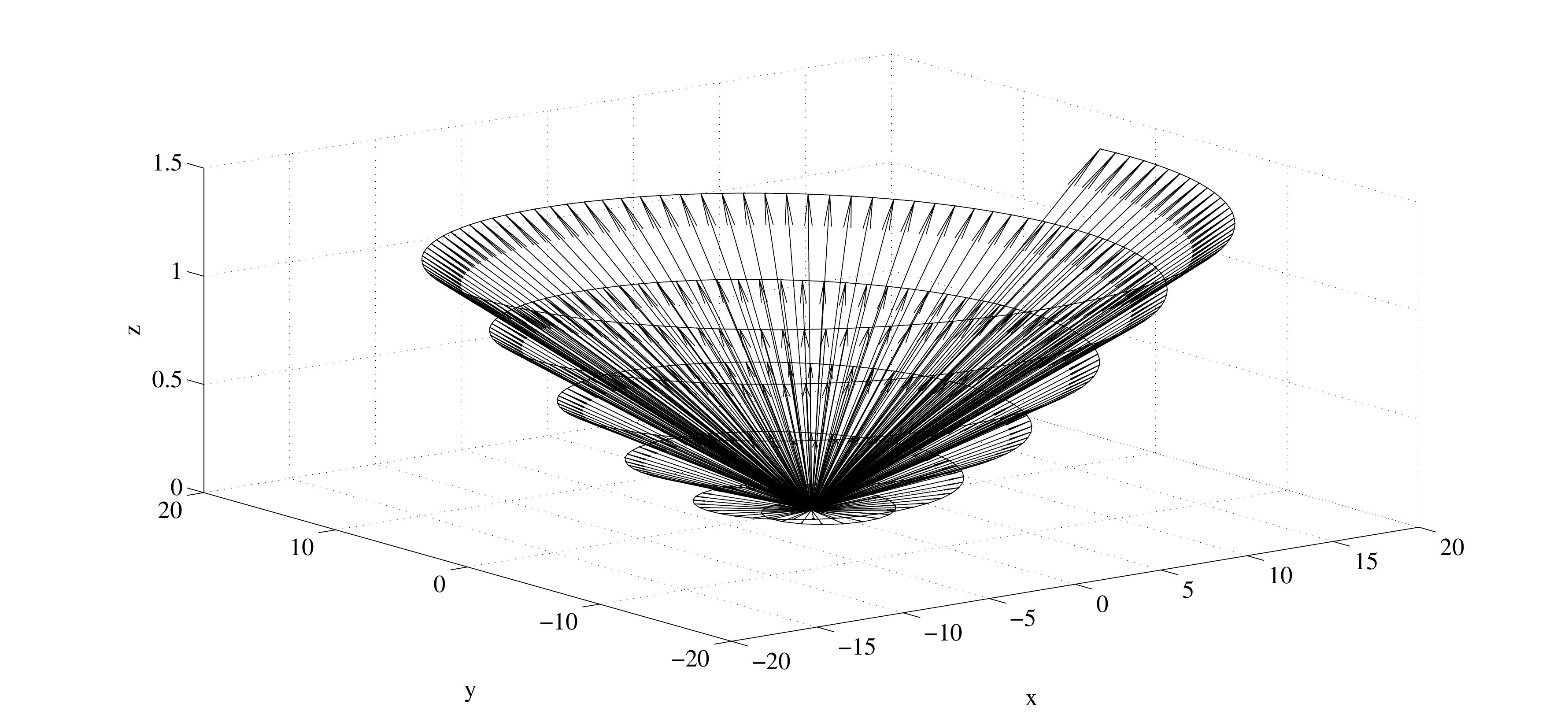}
\includegraphics[angle=0,width=1\textwidth]{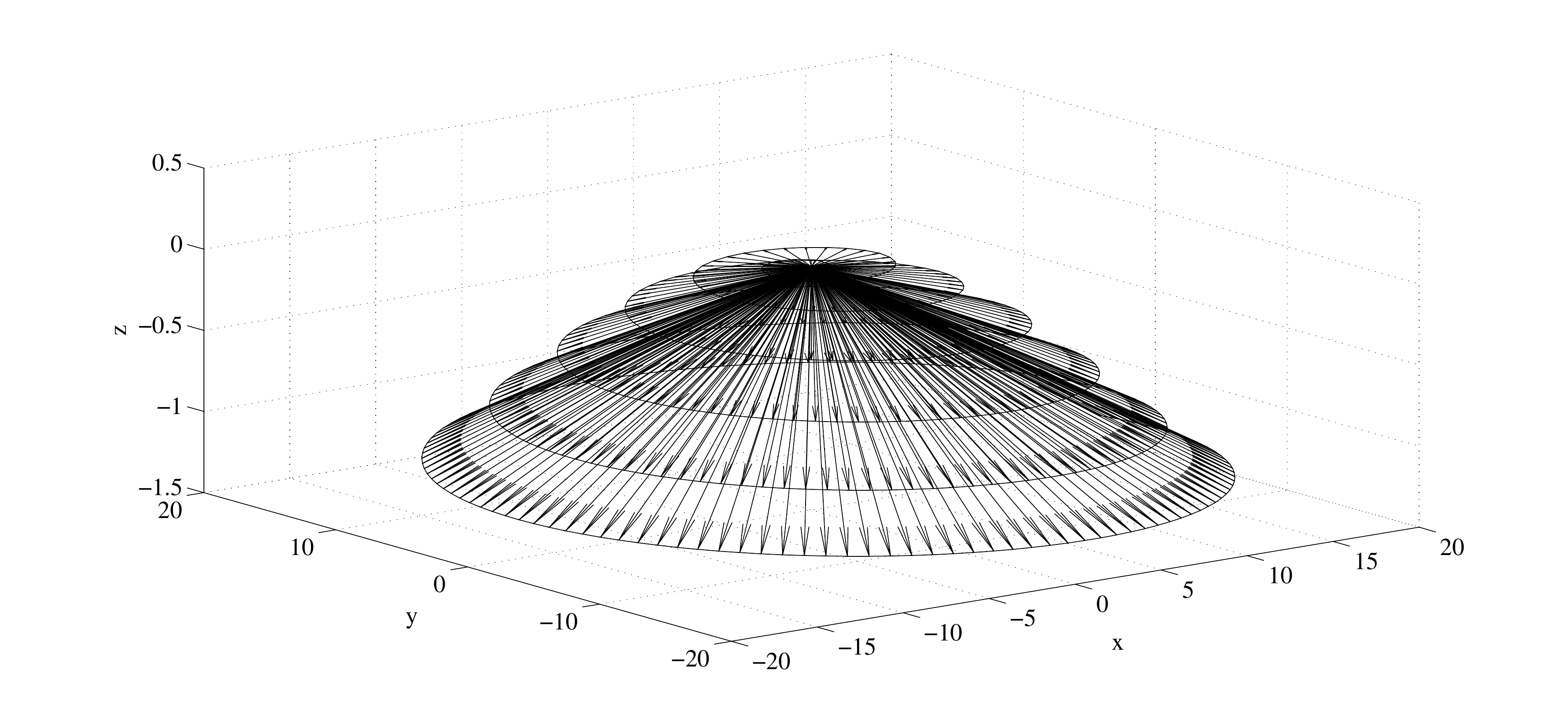}
\caption{Discrete spirals of Theodorus in 3D with $\left\lVert \mathbf{T}_{400} \right\rVert=\sqrt{400}$, angle between z-axis and $\mathbf{T}_{l}$, for $l\ge 19$, is automatically (a) decreasing (only once decreased by small amount, e.g. from $\frac{\pi}{2}$ to $(\frac{\pi}{2}-\frac{\pi}{720})$ radian (or \ang{90} to \ang{89.75}), at $\mathbf{T}_{18}$ for top figure) (b) increasing (only once increased by small amount, e.g. from $\frac{\pi}{2}$ to $(\frac{\pi}{2}+\frac{\pi}{720})$ radian (or \ang{90} to \ang{90.25}), at $\mathbf{T}_{18}$ for bottom figure) and they are drawn without any obstruction. \label{fig:dsot1}}
\end{figure}

This 2D discrete spiral of Theodorus can be extended to higher dimensions, using ND unit vectors ($\mathbf{x}_{l}$) such that it satisfy \eqref{dst1}, \eqref{dst3}, \eqref{dst31} and \eqref{dst4}. For example, two discrete spiral of Theodorus are constructed in 3D, as shown in Figure~\ref{fig:dsot1} with $\left\lVert \mathbf{T}_{400} \right\rVert=\sqrt{400}$, using 3D unit vectors.
It is interesting to note that, angle between positive z-axis and $\mathbf{T}_{l}$, for $l\ge 19$, is automatically (a) decreasing (only once decreased by small amount, e.g. from $\frac{\pi}{2}$ to $(\frac{\pi}{2}-\frac{\pi}{720})$ radian (or \ang{90} to \ang{89.75}), at $\mathbf{T}_{18}$ for top Figure~\ref{fig:dsot1}) (b) increasing (only once increased by small amount, e.g. from $\frac{\pi}{2}$ to $(\frac{\pi}{2}+\frac{\pi}{720})$ radian (or \ang{90} to \ang{90.25}), at $\mathbf{T}_{18}$ for bottom Figure~\ref{fig:dsot1}) and they are drawn without any obstruction or overlap in figure.
Here, we have obtained the spiral of Theodorus as a special case of energy preserving theorem (EPT), hence, in other word, we conclude that the EPT is a generalization of the spiral of Theodorus.
\subsection{FDM as a NTI system model}
Here, we prove that the FDM is NTI system model by counterexample.
 Let $x_1(t)=\sin(10\pi t)$ and $x_2(t)=\sin(100\pi t)$ in time interval $t\in[0,1]$ s. Clearly, both signal $x_1(t)$ and $x_2(t)$ are FIBFs, $FDM[x_1(t)]=x_1(t)$ and $FDM[x_2(t)]=x_2(t)$, thus $FDM[x_1(t)]+FDM[x_2(t)]=x_1(t)+x_2(t)$. However, it is very easy to verify, by the FDM algorithms, that $FDM[x_1(t)+x_2(t)]=\{x_{1}(t), x_{2}(t)\}$ which implies that $FDM[x_1(t)+x_2(t)]\ne FDM[x_1(t)]+FDM[x_2(t)]$. Clearly, FDM is a time-invariant system model as it follows~\eqref{nti0}, i.e. $FDM[x_1(t-\tau)]=x_1(t-\tau)$, $FDM[x_2(t-\tau)]=x_2(t-\tau)$ and $FDM[x_1(t-\tau)+x_2(t-\tau)]=\{x_{1}(t-\tau), x_{2}(t-\tau)\}$.
\subsection{Image decomposition}
\begin{figure}[!t]
\centering
\includegraphics[angle=0,width=1\textwidth]{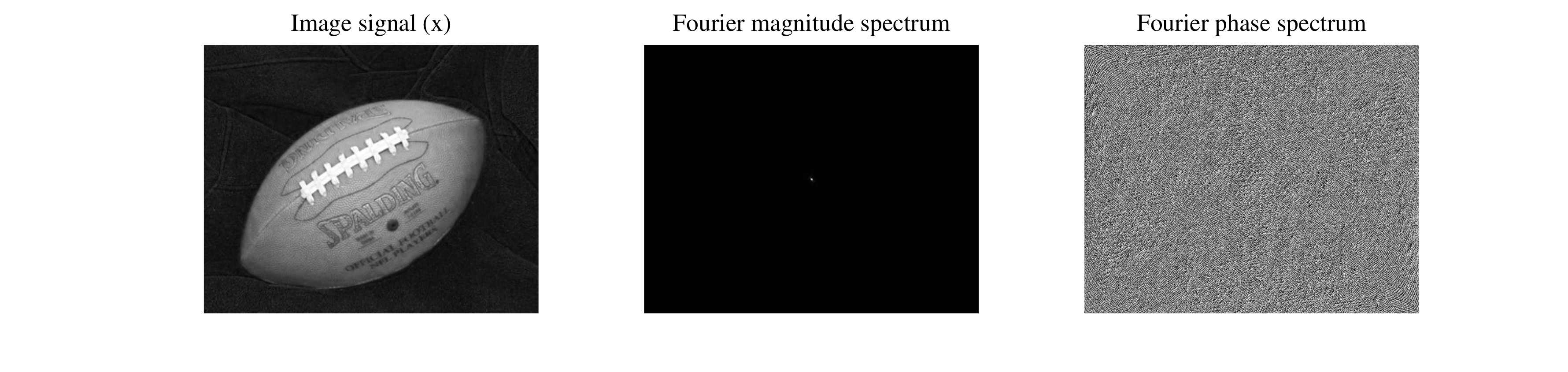}
\caption{Football image signal $\mathbf{x}$, its Fourier magnitude spectrum and phase spectrum. \label{fig:FootBall}}
\end{figure}

\begin{figure}[!t]
\centering
\includegraphics[angle=0,width=1\textwidth]{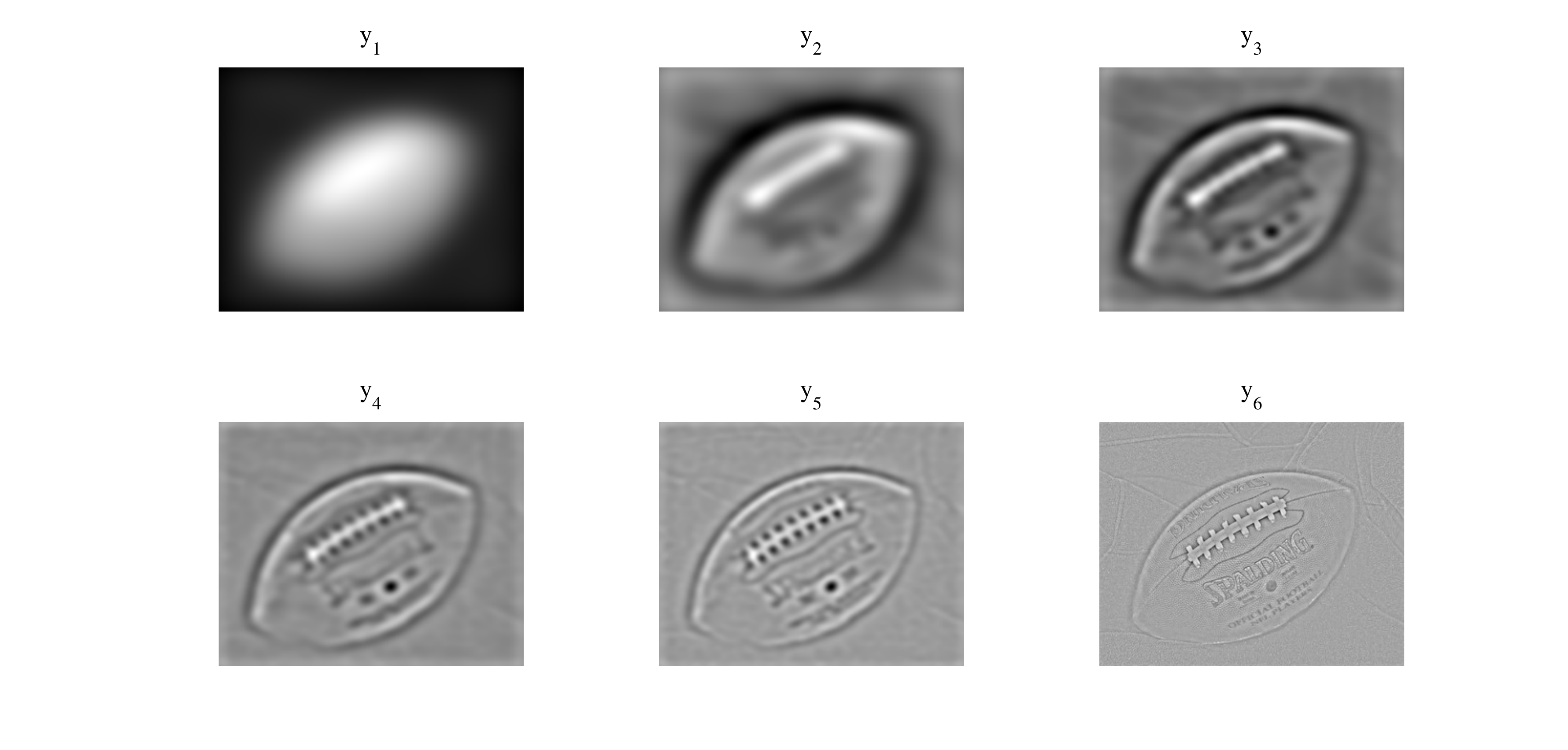}
\includegraphics[angle=0,width=1\textwidth]{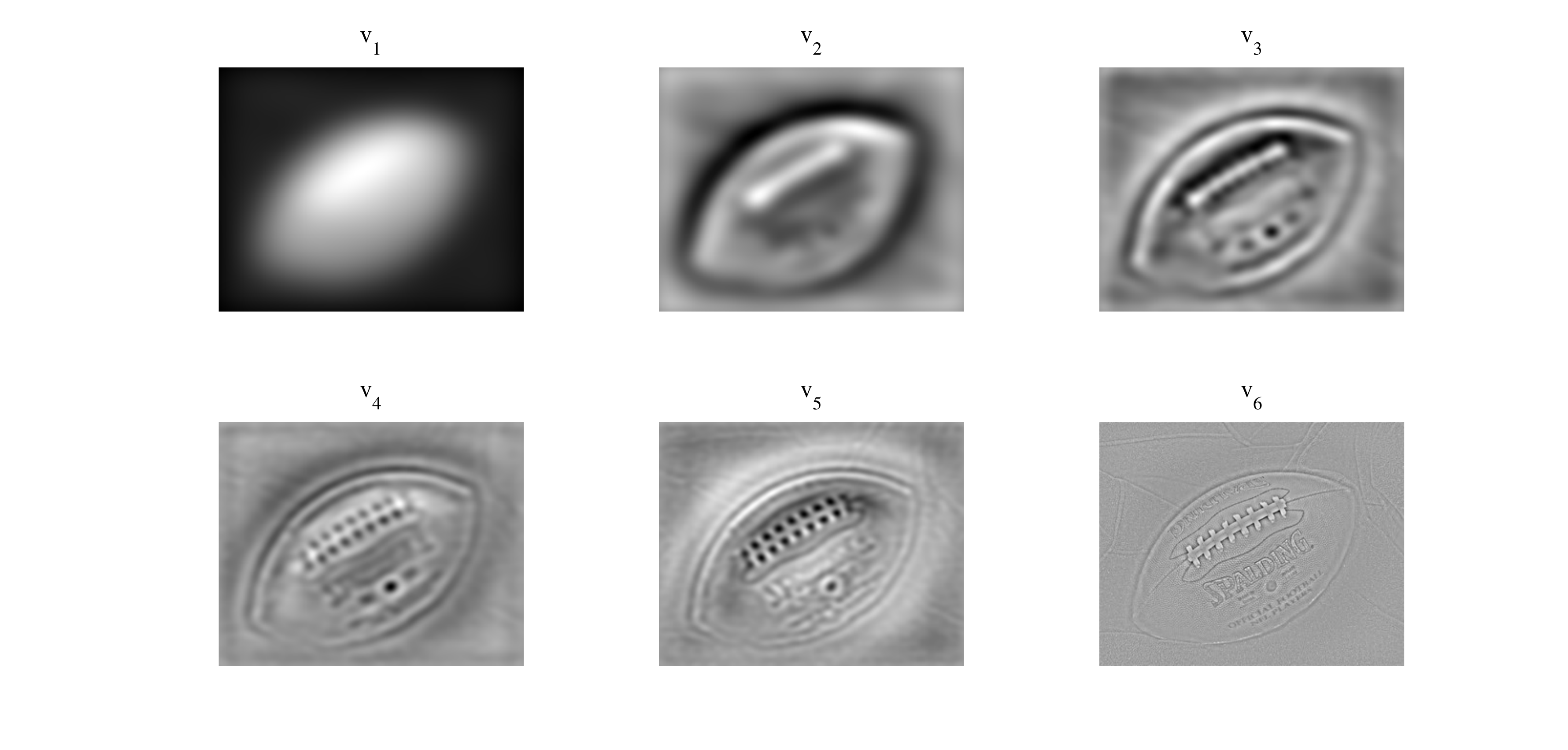}
\caption{A foot ball image $\mathbf{x}$ is decomposed by Gaussian low pass frequency domain filtering into: (a) $\{\mathbf{y}_i\}_{i=1}^6$ using Algorithm 1 (b) LINOEP vectors $\{\mathbf{v}_i\}_{i=1}^6$ using Algorithm 3.\label{fig:FootBall1}}
\end{figure}

\begin{figure}[!t]
\centering
\includegraphics[angle=0,width=1\textwidth]{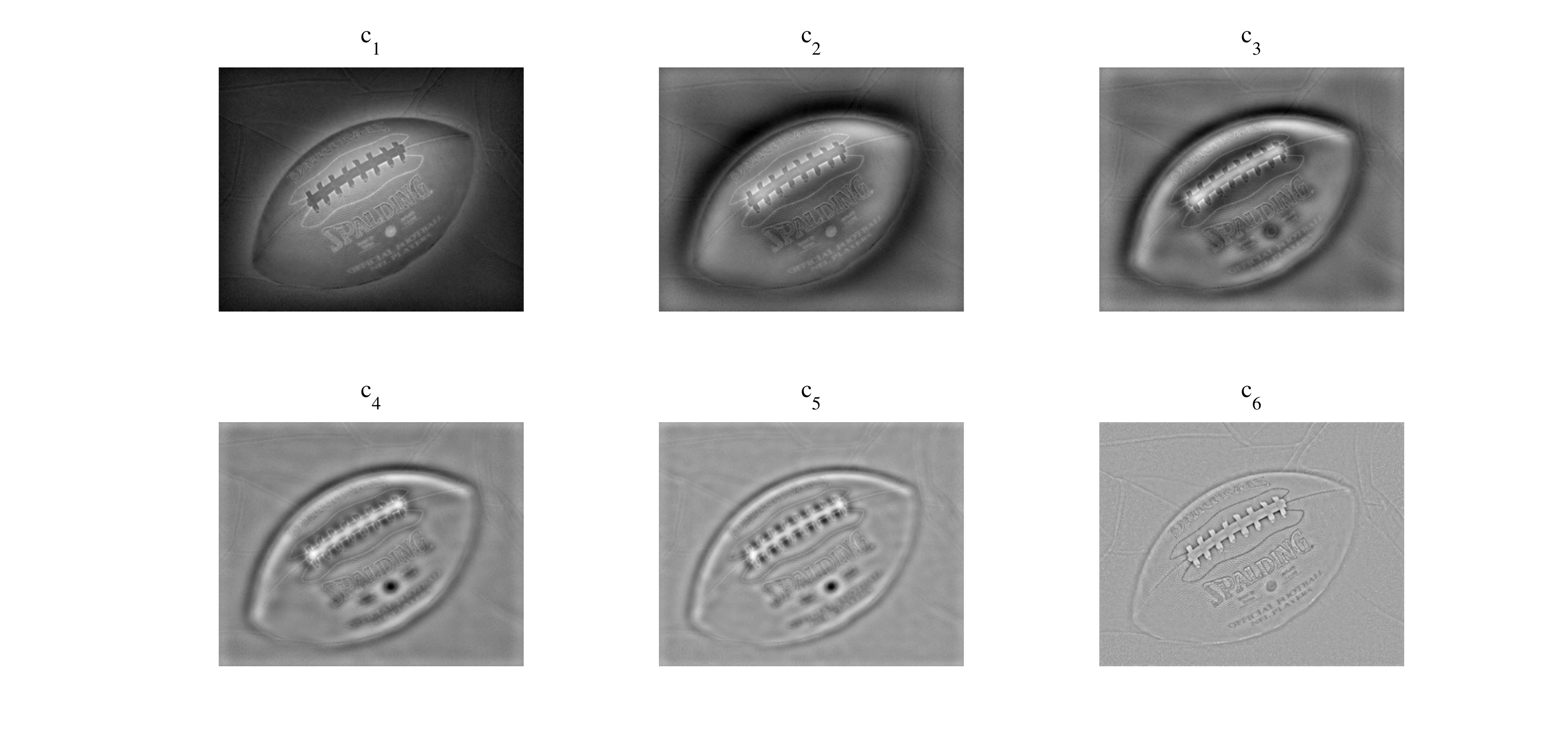}
\caption{A foot ball image $\mathbf{x}$ is decomposed by Gaussian low pass frequency domain filtering into LINOEP vectors $\{\mathbf{c}_i\}_{i=1}^6$ using Algorithm 2.\label{fig:FootBall2}}
\end{figure}

A Football image signal, $\mathbf{x}$, its Fourier magnitude spectrum and phase spectrum are shown in Figure~\ref{fig:FootBall}.
The energy computation of components generated from iterative frequency domain Gaussian low pass filtering, using Algorithm 1, Algorithm 2 and Algorithm 3, of football image are shown in Table~\ref{table:Etable}. The image components, $\{\mathbf{y}_i\}_{i=1}^6$, are obtained form from iterative frequency domain Gaussian low pass filtering using Algorithm 1 such that, $\mathbf{x}=\sum_{i=1}^6\mathbf{y}_i$, in Figure~\ref{fig:FootBall1}. Vectors, $\mathbf{y}_i$, are neither orthogonal nor LINOEP one, hence there is energy leakage and percentage error in energy (Pee) is 10\%, as calculated and shown in Table~\ref{table:Etable}. Vectors, $\mathbf{v}_i$ (Figure~\ref{fig:FootBall1}) and $\mathbf{c}_i$ (Figure~\ref{fig:FootBall2}), are obtained form iterative frequency domain Gaussian low pass filtering with orthogonalization in each iteration using Algorithm 3 and Algorithm 2, respectively. These two set of vectors are LINOEP one and hence there is no energy leakage and Pee is approximately zero as calculated and shown in Table~\ref{table:Etable}. There is not much visible difference between vectors $\{\mathbf{y}_i\}_{i=1}^6$ and $\{\mathbf{v}_i\}_{i=1}^6$ as shown in Figure~\ref{fig:FootBall1}. However, Figure~\ref{fig:FootBall2} is visibly different (better visual image quality) than Figure~\ref{fig:FootBall1}.

\begin{table}[!t]
\caption{The energy computation of components generated from iterative frequency domain Gaussian low pass filtering, Algorithm 1 and Algorithm 2. Energy of football image is $E_\mathbf{x}=665156949$.
Percentage error in energy=$\frac{(E_\mathbf{x}-\sum_{i=1}^6E_{\mathbf{x}_i})}{E_\mathbf{x}}\times100$.} 
\centering 
\begin{tabular}{|c|c|c|c|c|c|c|c|c|}
  \hline
   $i$  & 1 & 2 & 3 & 4 & 5 & 6 & $\sum_{i=1}^6E_{\mathbf{x}_i}$ & \% error \\\hline
   $E_{\mathbf{y}_i}$ &  $ 5.7088$ & $0.0760$ & $ 0.0167$ & $0.0068$ & $0.0038$ & $0.1080$ & $5.9201$ & $10.9967$ \\
   &  $\times 10^8$ & $\times 10^8$ & $\times 10^8$ & $\times 10^8$ & $\times 10^8$ & $\times 10^8$ & $\times 10^8$ & $\%$ \\\hline
   $E_{\mathbf{v}_i}$ &  $6.3321$ & $0.1399$ & $0.0424$ & $0.0224$ & $0.0156$ & $0.0991$ & $6.6516$ & $ 1.7922$ \\
   &  $\times 10^8$ & $\times 10^8$ & $\times 10^8$ & $\times 10^8$ & $\times 10^8$ & $\times 10^8$ & $\times 10^8$ & $\times 10^{-14}\%$ \\\hline
   $E_{\mathbf{c}_i}$ &  $5.4342$ & $0.2565$ & $0.0755$ & $0.0365$ & $0.0229$ & $ 0.8260$ & $6.6516$ & $ 1.7922$ \\
   &  $\times 10^8$ & $\times 10^8$ & $\times 10^8$ & $\times 10^8$ & $\times 10^8$ & $\times 10^8$ & $\times 10^8$ & $\times 10^{-14}\%$ \\
  \hline
\end{tabular}
\label{table:Etable} 
\end{table}
\section{Conclusion}
In this study, a general method is proposed to obtain a set of Linearly Independent Non-Orthogonal yet Energy (or square of the norm) Preserving (LINOEP) vectors using iterative filtering operation which we referred it as Filter Mode Decomposition (FDM).  We have shown that the general energy preserving theorem (EPT), which is valid for both linearly independent (orthogonal and nonorthogonal) and linearly dependent set of vectors, proposed by Singh P. et al. is a generalization of the discrete spiral of Theodorus (or square root spiral or Einstein spiral or Pythagorean spiral).
A novel class of vectors termed as `energy preserving vectors' are defined which can be a set of linearly independent or linearly dependent vectors.
We have shown that the (2D) discrete spiral of Theodorus is a special case of the EPT and multidimensional spirals can be easily obtained by the extension of 2D case, e.g. we have generated a 3D discrete spiral of Theodorus using a set of 3D energy preserving unit vectors.
We have also established that the recently proposed methods (e.g. Empirical Mode Decomposition (EMD), Variational Mode Decomposition (VMD), Eigenvalue Decomposition (EVD), Fourier Decomposition Method (FDM), etc.), for nonlinear and nonstationary time series analysis, are nonlinear time-invariant (NTI) system model of filtering. Simulation and numerical results, e.g. a decomposition of image into LINOEP components which are not only visually better quality images but also preserve energy, demonstrate the efficacy of LINOEP vectors.
\section*{Acknowledgment}
The author would like to thank JIIT Noida for permitting to carry out research at IIT Delhi and providing all required resources throughout this study.

\end{document}